\documentclass[prc,aps,superscriptaddress,twocolumn,showpacs,nofootinbib]{revtex4}

\usepackage[colorlinks,linkcolor=blue,anchorcolor=blue,citecolor=blue]{hyperref}
\usepackage[usenames,dvipsnames]{color}
\usepackage{graphicx,amsmath,amssymb,bm,multirow}
\usepackage{amsthm}
\usepackage{amsfonts}

\usepackage{dcolumn}
\usepackage{bm}

\usepackage[usenames,dvipsnames]{color}

\begin{document}

\title{Global study of beyond-mean-field correlation energies in covariant energy density functional theory using a collective Hamiltonian method}

\author{K. Q. Lu}
\affiliation{School of Physical Science and Technology, Southwest University, Chongqing 400715, China}

\author{Z. X. Li}
\affiliation{School of Physics and Nuclear Energy Engineering, Beihang University, Beijing 100191, China}

\author{Z. P. Li} \thanks{zpliphy@swu.edu.cn}
\affiliation{School of Physical Science and Technology, Southwest University, Chongqing 400715, China}

\author{J. M. Yao}
\affiliation{Department of Physics, Tohoku University, Sendai 980-8578, Japan}
\affiliation{School of Physical Science and Technology, Southwest University, Chongqing 400715, China}

\author{J. Meng} \thanks{mengj@pku.edu.cn}
\affiliation{State Key Laboratory of Nuclear Physics and Technology, School of Physics,
Peking University, Beijing 100871, China}
\affiliation{School of Physics and Nuclear Energy Engineering, Beihang University, Beijing 100191, China}
\affiliation{Department of Physics, University of Stellenbosch, Stellenbosch, South Africa}

\date{\today}

\begin{abstract}
We report the first global study of dynamic correlation energies (DCEs) associated with rotational motion and quadrupole shape vibrational motion in a covariant energy density functional (CEDF) for 575 even-even nuclei with proton numbers ranging from $Z=8$ to $Z=108$ by solving a five-dimensional collective Hamiltonian, the collective parameters of which are determined from triaxial relativistic mean-field plus BCS calculation using the PC-PK1 force. After taking into account these beyond mean-field DCEs, the root-mean-square (rms) deviation with respect to nuclear masses  is reduced significantly down to 1.14 MeV, which is smaller than those of other successful CEDFs: NL3* (2.96 MeV), DD-ME2 (2.39 MeV), DD-ME$\delta$ (2.29 MeV) and DD-PC1 (2.01 MeV). Moreover, the rms deviation for two-nucleon separation energies is reduced by $\sim34\%$ in comparison with cranking prescription. 
\end{abstract}

\maketitle


Nuclear energy density functional theory is nowadays one of the most important microscopic approaches for large-scale nuclear structure calculations in medium and heavy nuclei based on a universal energy density functional (EDF) with a few parameters constrained by the properties of finite nuclei and nuclear matter or neutron stars. It stands out as a unique microscopic model that can describe not only the masses of all existing nuclei, but also other key quantities for simulating nucleonthesis process, including beta decay rate and fission rate in a unified way~\cite{Bender03}. However, it is still a challenge for the current implementation of EDF to achieve satisfied accuracy. Therefore, great efforts have been devoted in many aspects to improving the accuracy of the EDF for atomic nuclei and subsequently deepening our understanding on the origin of elements in universe.

Nuclear binding energy or mass is one of the most fundamental properties of atomic nuclei.  
The root-mean-square (rms) deviation with respect to the measured nuclear masses in the EDF is typically around several MeV. Only after taking into account the beyond mean-field dynamic correlation energies (DCEs) in global fitting of the EDF to nuclear masses can one achieve the rms deviation of a few hundred keV~\cite{Gori09a,Gori13} by keeping a good description of nuclear matter properties. In these studies~\cite{Gori09a,Gori13}, the DCEs related to rotational and vibrational motions have been included phenomenologically with the {\em cranking prescription}. A better treatment of these DCEs is to carry out calculation with exact quantum number projection and generator coordinate method (GCM), which has been done based on a Skyrme SLy4  force~\cite{Bend05,Bend06} with the assumption of topological Gaussian overlap approximation (GOA) or based on Gogny D1M and D1S forces~\cite{Rodr14} without using the GOA.  In both the beyond mean-field (BMF) calculations, only axially deformed configurations are included and the obtained quadrupole DCEs are on average $3-4$ MeV. Although the BMF approaches with exact projections and GCM for triaxially deformed nuclei have already been developed in recent years~\cite{Bender08,Yao10,Rodriguez10}, they cannot be adopted for  a large-scale calculation of the DCEs because of extreme time-consuming. In contrast,  the method of five-dimensional collective Hamiltonian (5DCH) with parameters determined by the mean-field calculations is much cheaper in numerical realization and this method has been adopted to study low-lying states for a large set of even-even nuclei based on the Gogny D1S force~\cite{Dela10} and to optimize the new Gogny D1M force with a great success~\cite{Gori09b}.

Besides the nonrelativistic EDF, the covariant energy density functional (CEDF) has received wide attention and achieved great success in describing many phenomena of both stable and exotic nuclei~\cite{Reinhard89,Ring96,Vretenar05,Meng06}.  However, the description of nuclear binding energy based on the popular CEDFs is still not satisfactory.  It has been found in the most recent large-scale calculation of nuclear binding energy with axially deformed relativistic Hartree-Bogoliubov (RHB) for all proton number $Z\leq104$ even-even nuclei that the rms deviations with respect to the 640 measured masses of even-even nuclei in the AME2012 mass evaluation are 2.96 MeV for NL3*, 2.39 MeV for DD-ME2 and 2.29 MeV for DD-ME$\delta$, and 2.01 MeV for DD-PC1~\cite{Agbemava14}. Similar accuracy has been found in other global studies of nuclear masses with axially deformed RMF+BCS approaches~\cite{Lalazissis99,Geng05,Meng13}.

Therefore, a natural question to be asked is: to which extent the DCEs can improve the description of nuclear masses based on the existing CEDFs. To address this question,  we have recently carried out a global calculation of the DCEs for the 575 even-even nuclei with axially deformed relativistic mean-field plus BCS (RMF+BCS) approach using the point-coupling PC-PK1 parametrization~\cite{ZhangQS14}. The DCEs of deformed nuclei have been evaluated at the mean-field level with the cranking prescription adopted in building the mass tables based on Skyrme EDFs~\cite{Gori09a,Gori13}. After including these correlation energies, the rms deviation of the masses is reduced from 2.58 MeV to 1.24 MeV~\cite{ZhangQS14}. However, it should be pointed out that the cranking prescription for the DCE based on mean-field ground-state wave function is improper for the nuclei with transitional characters and for the nuclei with other energy minima competing with the ground-state one. Because the resultant DCE correction to binding energy in these cases depends much on the choice of the ground-state configuration which is, however, ill-defined and could be easily altered by the DCE~\cite{Fu13,ZhangQS14}. Moreover, for the transitional nuclei, large shape-fluctuation effect on nuclear properties is expected. All these deficiencies in the cranking prescription can in principle be taken care by the 5DCH method which is regarded as a GOA of angular momentum projection plus GCM. The 5DCH with the parameters determined by the CEDF has been implemented in Ref.~\cite{Niks09} and the success of it has been illustrated in a series of calculations for the spherical, transitional, and deformed nuclei from $A\sim40$ to superheavy region~\cite{Li09,Li11,Xiang12,Li12,Li13,Pras12,Fu13}. Recently, we have performed a detailed comparison between the 5DCH and exact projections plus GCM calculations for the low-lying states of $^{76}$Kr based on the PC-PK1 force and found that these two methods gave similar results~\cite{Yao14}.

In this work,  we revisit the DCEs in the CEDF using the 5DCH method for the 575 even-even nuclei with $8\leq Z \leq 108$ and examine the influence of these BMF correlation energies on the predicted masses and separation energies. Compared with our previous work on the DCEs based on the cranking prescription~\cite{ZhangQS14}, the present work takes into account more correlations in a more proper way as mentioned above and this paper presents the most advanced systematic calculation of the DCEs in the CEDF.

To this end, we first carry out a large-scale deformation constrained triaxial RMF+BCS calculation to generate the mean-field wave functions in the whole $(\beta, \gamma)$ plane, where the PC-PK1 force~\cite{Zhao10} in the particle-hole channel and a separable paring force~\cite{Tian09,Dug04,Dug08,Les09} in the particle-particle channel are adopted. This separable pairing force has finite range and preserves translational invariance. In particular, the numerical calculation is much simpler than that of Gogny force, which makes it more suitable for systematic studies. Several studies have already shown that the separable pairing force is quite successful in the description of nuclear shape transition and low-lying spectrum~\cite{Xiang12,Li12,Fu13}. The Dirac equation for single-particle wave function is solved by expanding in a set of eigenfunctions of 3D harmonic oscillator in cartesian coordinate with 12, 14, and 16 major shells for nuclei with $Z<20$, $20\leq Z<82$, and $Z\geq82$, respectively. The single-particle wave functions, occupation probabilities, and quasiparticle energies are used to calculate the mass parameters, moments of inertia, and collective potentials in the 5DCH, all of which are functions of the deformations $\beta$ and $\gamma$. The properties of nuclear ground state $0^+_1$ and other low-lying excited states are obtained by solving the 5DCH~\cite{Niks09,Li09}. The DCE $E_{\rm corr}$ is defined as the energy difference between the mean-field ground state $E^{\rm MF}_{\rm g.s.}$ and the 5DCH ground state $E^{\rm 5DCH}_{0^+_1}$. Considering the facts that the PC-PK1 force was parameterized to many spherical nuclei ranging from O to Pb and the 5DCH method for nuclei around double-closed shell is improper, these lead to erroneous negative DCEs. For these nuclei, the $E_{\rm corr}$ is imposed to be zero in this work.


\begin{figure}[htb]
\includegraphics[width=8.5cm]{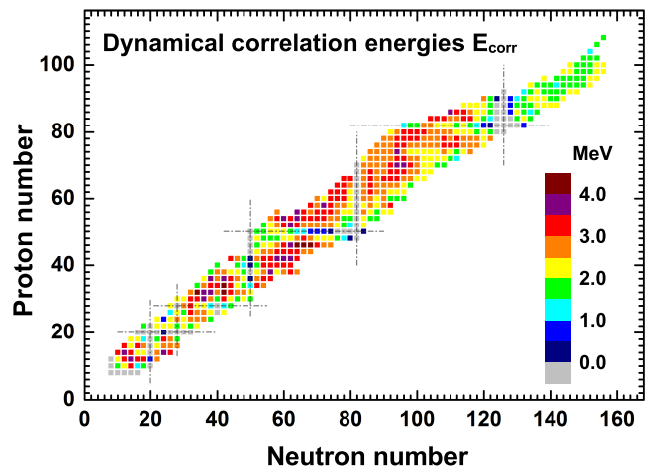}
\caption{\label{fig1}(Color online) Contour map of the quadrupole dynamical correlation energies
$E_{\rm corr}$ calculated by the CEDF based 5DCH model as functions of the neutron and proton numbers.}
\end{figure}

Figure \ref{fig1} displays the contour map of the quadrupole dynamical correlation energies $E_{\rm corr}$ calculated by 5DCH based on CEDF with PC-PK1 force. The quadrupole dynamical correlation energies range from 0 to $\sim4.4$ MeV, and mostly vary in the region of $2.0\sim3.5$ MeV. For the semi-magic nuclei with $Z=28, 50, 82$ and $N=28$, the correlation energies are nonzero  or even rather large. This is because the energy surfaces of these nuclei are either soft around energy minimum or with shape coexisting phenomena. The $E_{\rm corr}$ of the transitional nuclei with $Z\sim54, 78$ and $N\sim60, 90$ is pronounced, generally larger than $2.5$ MeV due to the shape fluctuation. The $E_{\rm corr}$ of the well-deformed nuclei with $A\sim170$ is around 2.5 MeV and it reduces to $\sim2.0$ MeV for the heavier nuclei with $A\sim250$.

\begin{figure}[htb]
\includegraphics[width=8cm]{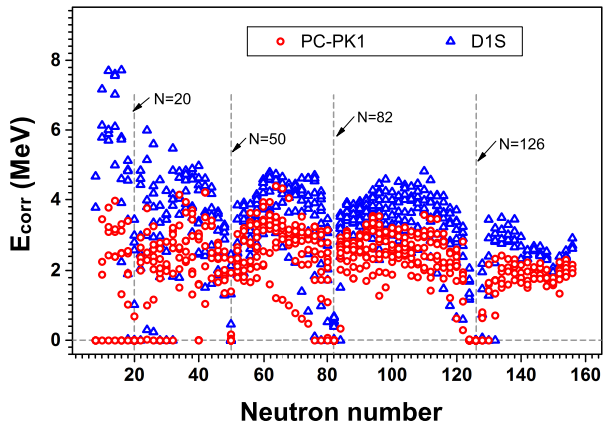}
\caption{\label{fig2}(Color online) The quadrupole dynamical correlation energies calculated
by 5DCH based on CEDF with PC-PK1 force (open circles) in comparison with those
based on HFB using Gogny D1S force (upper triangles).}
\end{figure}

In Fig.~\ref{fig2}, we compare our dynamical correlation energies with those calculated from 5DCH model based on Hartree-Fock-Bogoliubov (HFB) calculation using Gogny D1S force~\cite{Dela10}. The systematics of $E_{\rm corr}$ are similar in both calculations. However, the DCEs from PC-PK1 force are systematically smaller than those from Gogny D1S force, and the rms deviation of these two results is $\sim 1.95$ MeV. The difference in the DCE might be originated from the collective parameters (especially the zero point energies), which are sensitive to the effective interactions, in particular, to the pairing properties~\cite{Li11b}.

\begin{figure}[htb]
\includegraphics[width=7cm]{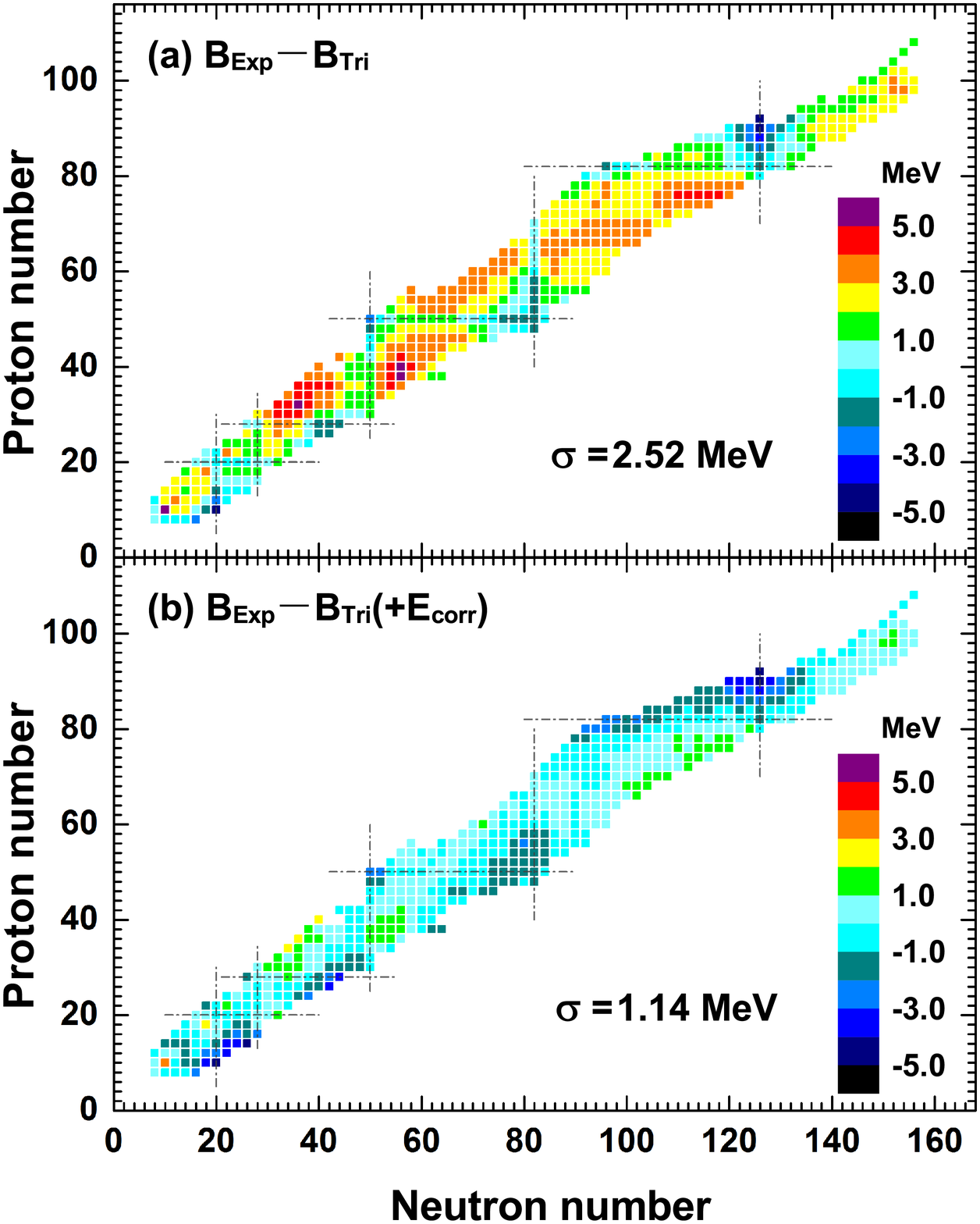}
\caption{\label{fig3}(Color online) Discrepancy of the CEDF calculated binding energies by PC-PK1
with the data for  575 even-even nuclei~\cite{audi12}. In panel (a), the CEDF calculated binding energies
are given by the binding energies of the lowest mean-field states, while in panel (b)
the dynamical correlation energies are taken into account.}
\end{figure}

Figure~\ref{fig3} displays the discrepancy of the calculated binding energies with respect to the data for 575 even-even nuclei~\cite{audi12}. The mean-field binding energy (denoted as $B_{\rm Tri}$) is obtained by adding the energy correction from static triaxial deformation to that from our previous axially deformed RMF+BCS calculations~\cite{ZhangQS14}. Such energy correction is defined as the energy difference between the ground state and the lowest non-triaxial minimum of the triaxial deformed nucleus (totally 31 in the present work) calculated by the triaxial RMF+BCS with the same inputs as in Ref.~\cite{ZhangQS14}. The static triaxial deformation effect makes the nucleus more binding in an average value of $\sim0.36$ MeV, closer to the experimental data.
Our main findings are as follows:
\begin{enumerate}
 \item[(i)]  The mean-field binding energies are systematically underestimated by about 3 MeV.  By including the DCEs from the 5DCH calculation as shown in Fig.~\ref{fig1}, the rms deviation is reduced from 2.52 MeV to 1.14 MeV, which is about 100 keV improvement in comparison with the cranking prescription in Ref.~\cite{ZhangQS14}. The remaining discrepancy between the calculated and measured  binding energies for most nuclei is in between $-1.0$ and 1.0 MeV.
 \item[(ii)] The large discrepancy for nuclei in the neighborhood of $(Z, N)\sim(40, 40)$
                in the cranking approximation calculation~\cite{ZhangQS14}
                has been modified significantly in the present calculation.
                This is because the shape coexistence and large shape fluctuation
                in this mass region are considered in 5DCH method.
 \item[(iii)] Similar as in Skyrme-HFB~\cite{Gori09a,Gori13} and Gogny-HFB~\cite{Gori09b} mass models, large deviations occur around magic numbers. In particular, the binding energies of $N\sim126$ isotones are significantly overestimated, which has also been found in other CEDFs~\cite{Agbemava14}. It is still an open question which has not been understood.
\end{enumerate}

\begin{figure}[htb]
\includegraphics[width=7cm]{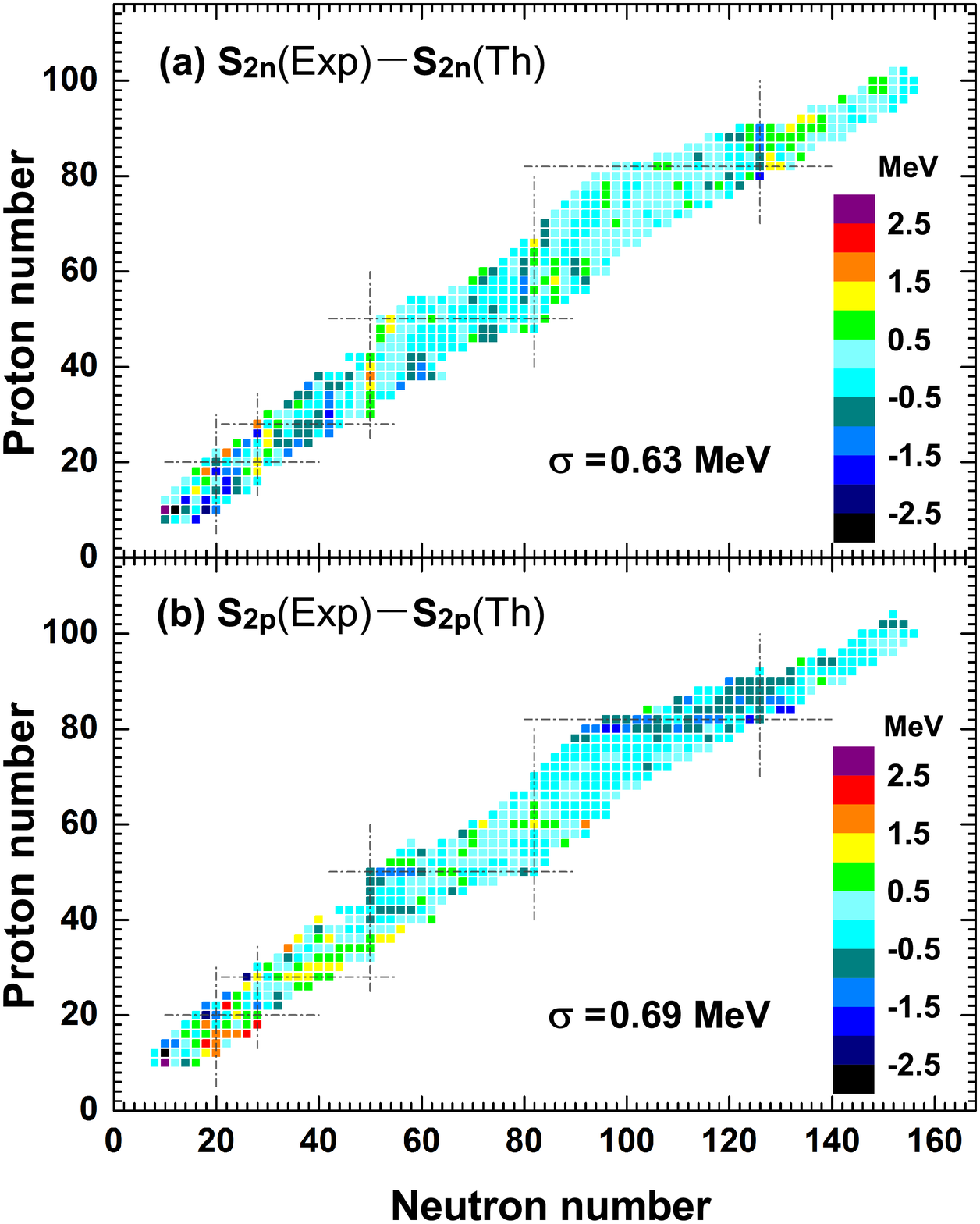}
\caption{\label{fig4}(Color online) Discrepancy of the theoretical two-neutron (panel a)
and two-proton (panel b) separation energies extracted from the calculated binding energies including the DCEs
with respect to the data~\cite{audi12}.}
\end{figure}

Figure~\ref{fig4} shows the discrepancy of the CEDF calculated two-neutron and two-proton separation energies by PC-PK1 with respect to the data~\cite{audi12}. It is remarkable that the separation energies are reproduced very well except for the nuclei around $N\sim20, 126$ and $Z\sim28, 82$. The discrepancy between the calculated and measured values for most nuclei is in between -0.5 and 0.5 MeV, and the rms deviations are 0.62 MeV and 0.69 MeV for $S_{2n}$ and $S_{2p}$, respectively, which are smaller than the values 0.96 MeV and 1.03 MeV  in Ref.~\cite{ZhangQS14} based on the cranking prescription.   For comparison, we also compute the separation energies from the binding energies of the lowest mean-field states (i.e. $B_{\rm Tri}$ in Fig.~\ref{fig3}) and the rms deviations are respectively 0.91 MeV and 0.96 MeV for neutrons and protons.  In short, the DCEs from the 5DCH calculation gives a significant improvement for the description of two-nucleon separation energies.


In summary, we have carried out a global study of the dynamical correlation energies associated with the quadrupole shapes for the 575 even-even nuclei with proton numbers ranging from $Z=8$ to $Z=108$  by solving the 5DCH with the collective parameters determined from the CEDF calculation using PC-PK1 force. The DCEs range from 0 to $\sim4.4$ MeV, and mostly vary in the region of $2.0\sim3.5$ MeV. After including DCEs, the CEDF predictions for 575 masses, 521 two-neutron separation energies, and 497 two-proton separation energies are improved significantly with the rms deviations 1.14 MeV, 0.62 MeV, and 0.69 MeV, respectively. It is remarkable that the rms deviation for two-nucleon separation energies is reduced by $\sim34\%$ in comparison with our previous results based on cranking prescription.

This work was supported in part by National Undergraduate Training Programs for Innovation and Entrepreneurship (Project No.201310635059), the Major State 973 Program 2013CB834400, the NSFC under Grants No. 10975008, No. 11305134, No. 11175002, No. 11105110, No. 11105111, No. 11335002, and No. 11475140, the Research Fund for the Doctoral Program of Higher Education under Grant No. 20110001110087.



\end{document}